\documentclass[twocolumn, switch]{article}

\usepackage{preprint}
\usepackage{algorithm}
\usepackage{algorithmic}
\usepackage{amsmath, amsthm, amssymb, amsfonts}
\usepackage{lineno}
\usepackage[numbers,square]{natbib}

\usepackage[utf8]{inputenc}
\usepackage[T1]{fontenc}
\usepackage{xcolor}
\usepackage[colorlinks = true,
            linkcolor = blue,
            urlcolor  = blue,
            citecolor = blue,
            anchorcolor = blue]{hyperref}
\usepackage{booktabs}
\usepackage{nicefrac}
\usepackage{microtype}
\usepackage{lineno}
\usepackage{float}
\usepackage{lipsum}

\usepackage{newfloat}
\DeclareFloatingEnvironment[name={Supplementary Figure}]{suppfigure}
\usepackage{sidecap}
\sidecaptionvpos{figure}{c}
\usepackage{graphicx}
\usepackage{subcaption}

\usepackage{titlesec}
\titlespacing\section{0pt}{12pt plus 3pt minus 3pt}{1pt plus 1pt minus 1pt}
\titlespacing\subsection{0pt}{10pt plus 3pt minus 3pt}{1pt plus 1pt minus 1pt}
\titlespacing\subsubsection{0pt}{8pt plus 3pt minus 3pt}{1pt plus 1pt minus 1pt}

\title{Emergent Active Clusters through Wave-Mediated Interactions}

\usepackage{eso-pic}
\usepackage{tikz}
\usepackage{xcolor}
\PassOptionsToPackage{colorlinks=true,linkcolor=gray,urlcolor=gray}{hyperref}

\newcommand{\AddMyWatermarks}{%
  \begin{tikzpicture}[remember picture, overlay]
    \node[color=gray!90, scale=1] at ([xshift=0in,yshift=-5in]current page.center) {%
    };
  \end{tikzpicture}%
}

\AddToShipoutPictureBG*{\AddMyWatermarks}

\usepackage{titling}
\usepackage{orcidlink}
\usepackage{footmisc}
\newcommand{\Author}[3]{% Name, ORCID, Institution
  \textbf{#1}\textsuperscript{#2},\ \orcidlink{#3} %
}

\author{
  \Author{Yuanmei Li}{1}{0009-0006-8799-175X} \and
  \Author{Rahil Valani}{2}{0000-0001-8346-0739}
}

\date{%
  \textsuperscript{1}School of Physics, Nanjing University, 22 Hankou Road, Gulou District, Nanjing, Jiangsu 210093, China\\
  \textsuperscript{2}Rudolf Peierls Centre for Theoretical Physics, Parks Road, University of Oxford, OX1 3PU, United Kingdom\\[1em]
  \footnotesize \textbf{Corresponding author:} Yuanmei Li \texttt{<lymphy@smail.nju.edu.cn>}\\
  Rahil Valani \texttt{<rahil.valani@physics.ox.ac.uk>}\\
}

\begin{document}

\twocolumn[{
  \begin{@twocolumnfalse}

\maketitle
\thispagestyle{empty}

\begin{abstract}
Collective behaviour in active matter usually emerges from interactions between individually stable units. Here we demonstrate a different route to collective organization using droplets on a vibrated fluid bath. Large droplets that are unstable in isolation become persistent active clusters through a shared Faraday wavefield. The same wave-mediated interactions that stabilize these clusters also govern their collective dynamics, producing stationary, rotating, and translating states as droplet size is varied. Our results show how field-mediated interactions can transform individually unstable particles into coherent active entities with emergent structure and function.
\end{abstract}

%\keywords{Dynamical bonds \and Active matter \and Self-assembly \and Wave-mediated interactions \and Programmable materials}
%\vspace{0.35cm}

\vspace{0.35cm}

  \end{@twocolumnfalse}
}]

\section*{\textbf{Introduction}}

One of the central themes of physics is that interactions between simple constituents can give rise to emergent objects whose properties cannot be understood from those of the individual components alone~\citep{doi:10.1126/science.177.4047.393,doi:10.1126/science.284.5411.87}. Examples range from quasiparticles in condensed matter and coherent structures in fluids to collective states in active matter, where interactions generate new dynamical entities with behaviours absent at the level of isolated particles. Understanding how such emergent objects form, persist, and acquire their own dynamics remains a central challenge across many nonequilibrium systems.

%Active matter provides a particularly rich setting for studying how collective interactions generate emergent dynamical objects. Local interactions between self-driven particles give rise to collective states spanning flocking and swarming~\citep{Vicsek2012Collective}, motility-induced phase separation~\citep{Cates2015}, and active turbulence~\citep{Alert2022}. While much attention has focused on the collective phases that arise from these interactions, considerably less is understood about how interactions themselves create new coherent entities with their own structure and dynamics. An important class of active systems is distinguished by interactions that are mediated through a shared field such as hydrodynamic flows~\citep{Marchetti2013Soft}, chemical gradients~\citep{Hokmabad2022Chemotactic}, and optical fields~\citep{Wang2021Robot}, rather than through direct particle-particle forces. In these systems, particles continuously generate and respond to the surrounding field, creating feedback between the constituents and the medium that can drive self-organization and collective dynamics. Understanding how such field-mediated interactions give rise to coherent active structures remains an important open question in nonequilibrium physics.

Active matter provides a particularly rich setting for studying how collective interactions generate emergent dynamical objects. Local interactions between self-driven particles give rise to collective states spanning flocking and swarming~\citep{Vicsek2012Collective}, motility-induced phase separation~\citep{Cates2015}, and active turbulence~\citep{Alert2022}. Among these, an important class of systems is distinguished by interactions that are mediated through a shared field, such as hydrodynamic flows~\citep{Marchetti2013Soft}, chemical gradients~\citep{Hokmabad2022Chemotactic}, or optical fields~\citep{Wang2021Robot}, rather than through direct particle-particle forces. In these systems, particles continuously generate and respond to the surrounding field, creating a feedback loop that naturally couples the dynamics of the entire assembly. How such field-mediated interactions give rise to coherent active objects with persistent structure and collective motion remains largely unexplored.

In this work, we use droplets on a vibrating fluid bath as a minimal realization of a field-mediated active system. Such droplets interact through a long-lived Faraday wavefield that is continuously generated and sampled by the droplets themselves, establishing a feedback loop between particle motion and the surrounding field~\citep{Couder2005Nature}. We identify a distinct regime of wave-mediated active clustering in which droplets that are unstable in isolation assemble into persistent active clusters through this shared wavefield. The same wave-mediated interactions that stabilize these clusters also govern their collective dynamics, giving rise to stationary, rotating and translating active states selected by droplet size and driving strength. These clusters therefore behave as coherent dynamical objects whose morphology and motion emerge collectively from interactions encoded in the shared wavefield, illustrating how field-mediated interactions can transform simple, unstable particles into stable, active entities with emergent structure and dynamics.

\section*{\textbf{Results}}

\subsection*{\textbf{Emergence of Active Clusters}}

\begin{figure*}[t]
  \centering
  \includegraphics[width=\linewidth]{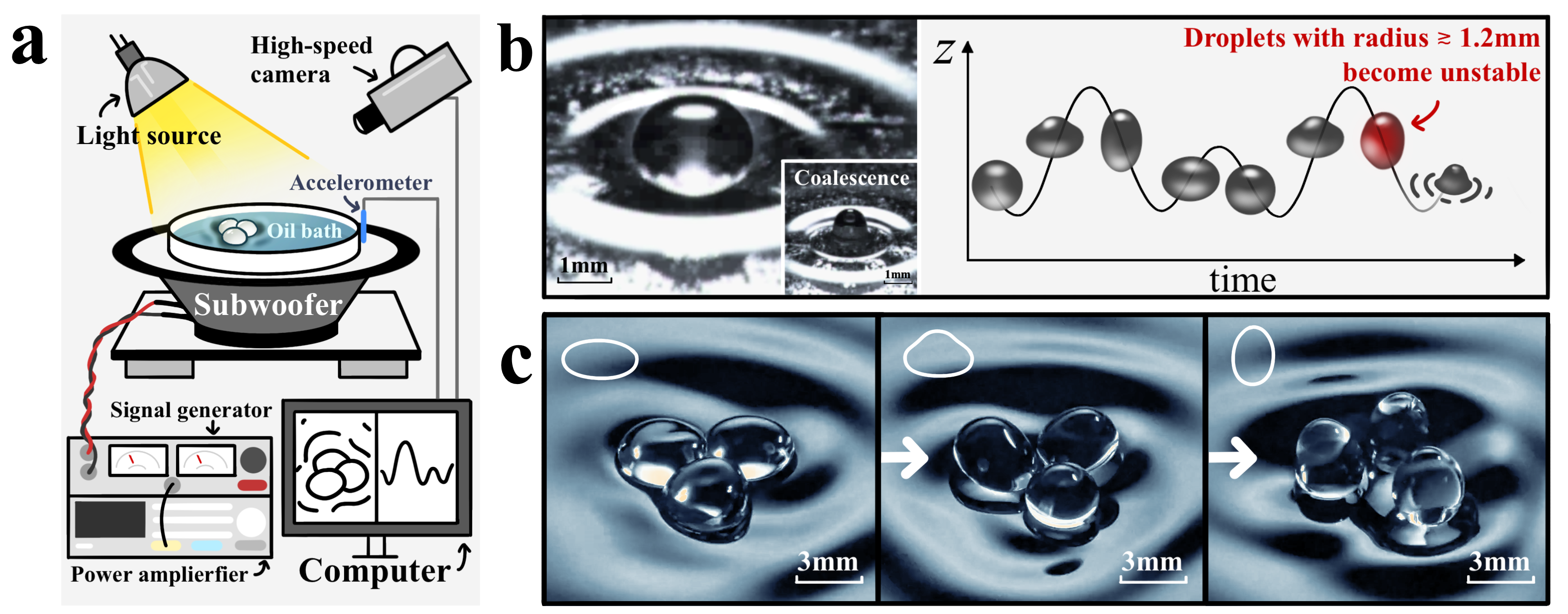}
\caption{\textbf{Emergence of active clusters from individually unstable droplets.}
\textbf{(a)} Experimental setup. A circular fluid bath is driven by dual-frequency forcing (80\,Hz and 40\,Hz), while droplet motion is recorded using a high-speed camera.
\textbf{(b)} Instability of an isolated large droplet ($R\gtrsim1.2$\,mm). The droplet loses phase synchronization with the vibrating bath and rapidly coalesces with the liquid surface (see Movie~1).
\textbf{(c)} Synchronized bouncing within an active cluster. Three large droplets bounce in synchrony over an entire vibration cycle. This synchronous shape deformation is also clearly observed in clusters of 2 to 6 droplets (see Movie~2). 
}
\label{Fig1}
\end{figure*}

Walking droplets on a vertically vibrated fluid bath provide a well-established model system in which particles interact through a self-generated, memory-bearing wavefield \cite{Walker1978,Couder2005Nature,Valani2019Superwalking,Eddi2011}. In this system, a droplet repeatedly impacts the liquid surface, exciting localized subharmonic Faraday waves~\citep{Faraday1831a} whose gradual decay imparts memory to the interface. The droplet subsequently is guided horizontally by the local wave slope from its self-generated wavefield, establishing a feedback loop between particle and field and forming a stable, self-propelled, wave-particle entity on the liquid surface. Owing to the wave-particle coupling, remarkably, these walking droplets exhibit several hydrodynamic quantum analogs~\citep{Bush_2020,Bush2024}. Through this wave-mediated interaction, droplets can form a remarkable spectrum of bound states, including orbiting pairs~\citep{Oza2017Orbiting,Protiere2006PWA,Protière2008}, promenading pairs~\citep{Arbelaiz2018,Borghesi2014}, chasing pairs~\citep{Valani2019Superwalking}, ratchets~\citep{GaleanoRios2018,Eddi2008}, and self-organized droplet lattices and small spinning clusters of droplets~\citep{lieber2007,Eddi2009Lattice,Protiere2005,Valani2019Superwalking}. Moreover, solid floating bodies at air-fluid interface, when vibrated internally or externally, can also form bound states and self-organized structures through wave-mediated interactions~\citep{353x-p2dx,PhysRevE.111.035104,Thomson2023,PhysRevFluids.8.L112001}

Under single-frequency forcing, walking droplets are typically small (radii $\sim0.3$--$0.5$\,mm) and slow (up to $15$\,mm/s)~\citep{Molacek2013}. When the bath is driven simultaneously at two frequencies, a primary frequency and its subharmonic, droplets enter the superwalking regime, in which stable walking is sustained for significantly larger droplets (radii up to $\sim1.2$\,mm) that move faster (up to $\sim50$\,mm/s), owing to enhanced vertical synchronization with the bath vibrations~\citep{Valani2019Superwalking,Pranav2024}. As the droplet size approaches the upper limit of this regime, the vertical dynamics change qualitatively: instead of undergoing pronounced bounce cycles, the droplets spend an increasing fraction of each forcing cycle in contact with the bath, exhibiting large periodic shape deformations with only weak lift-off.

We perform experiments on a circular bath driven simultaneously at $80$\,Hz and $40$\,Hz using a subwoofer-based shaking system (Fig.~\ref{Fig1}a; Methods). Extending beyond the conventional superwalking regime, we find that droplets larger than approximately $1.2$\,mm can no longer sustain stable synchronized shape deformations under the dual-frequency forcing used here and the droplets rapidly coalesce with the liquid surface (Fig.~\ref{Fig1}b; Movie~1). Remarkably, this behaviour changes in the collective setting. When several such large droplets are generated in close proximity, they no longer coalesce. Instead, their deformation cycles become synchronized and the droplets spontaneously assemble into long-lived, compact active clusters (Fig.~\ref{Fig1}c). Throughout this work, we define an \emph{active cluster} as a group of droplets that maintain a persistent collective configuration. The emergence of these clusters demonstrates that collective wave-mediated interactions stabilize a coherent dynamical object that cannot exist at the level of isolated droplets.

\subsection*{\textbf{Wave-Mediated Cluster Formation}}

\begin{figure*}[ht]
  \centering
  \includegraphics[width=\linewidth]{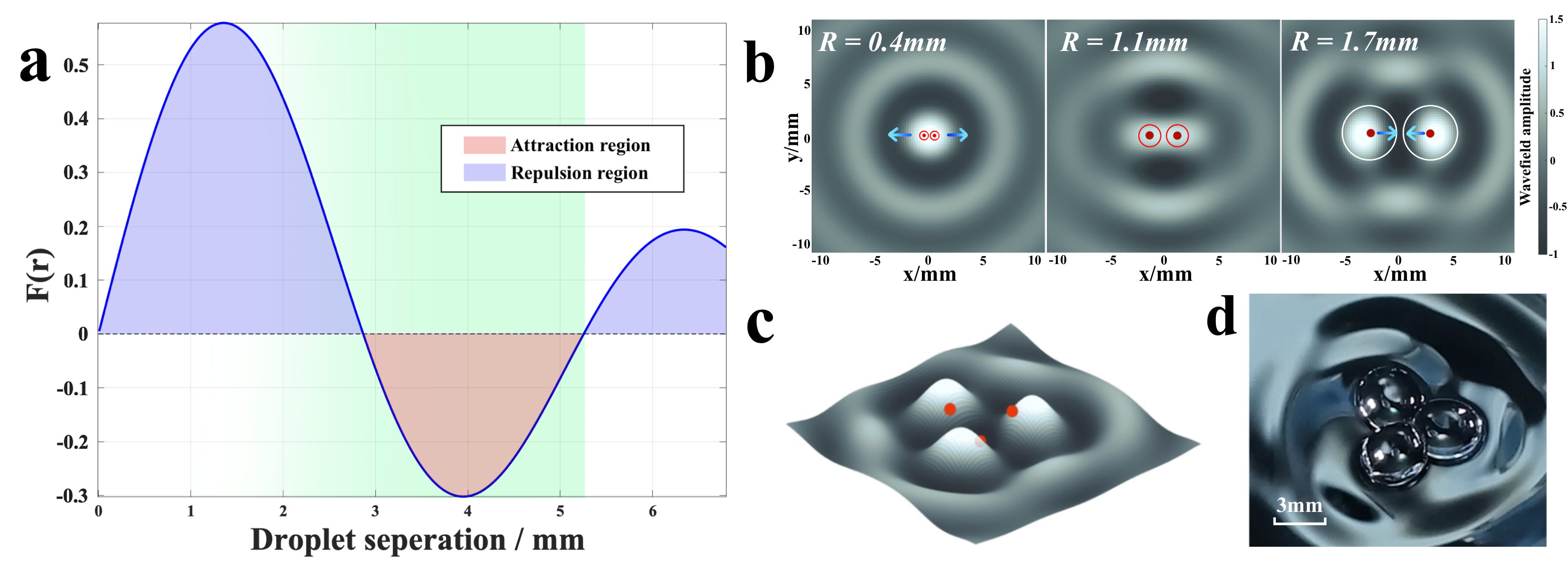}
\caption{\textbf{Wave-mediated interactions underlying active cluster formation.}
\textbf{(a)} Effective interaction force profile $F(r)$ obtained from the superposition of two droplet-generated wavefields. The zero crossing defines the equilibrium spacing between neighbouring droplets.
\textbf{(b)} Evolution of the wave-mediated interaction with droplet size. Superimposed strobed wavefield from Eq.~\eqref{eq: strobe} showing interference between two droplets produces predominantly repulsive interactions for small walkers ($R=0.4$\,mm), weak interactions for intermediate superwalkers ($R=1.1$\,mm), and strong short-range attraction for larger droplets ($R=1.7$\,mm), enabling stable active clusters.
\textbf{(c)} Wavefield landscape obtained by superposing the strobed wavefields for three large droplets. The resulting interaction landscape illustrates how wavefield superposition naturally extends to multi-droplet configurations.
\textbf{(d)} Experimental wavefield of the corresponding three-droplet cluster.
}
\label{Fig2}
\end{figure*}

The active clusters observed above are mediated by the shared Faraday wavefield generated by the droplets themselves. Each bouncing droplet continuously excites localized standing waves that interfere with those of neighbouring droplets, producing a dynamic interaction landscape on the liquid surface. Droplets respond to the local gradient of this evolving wavefield, so that the effective interactions are determined by collective wave interference. This shared wavefield therefore provides the physical mechanism that maintains both the structure and dynamics of the active clusters.

To describe the wave-mediated interactions, we model the wavefield generated by each droplet using the axisymmetric strobed response of a bouncing droplet on a vibrating bath~\citep{Thompson2020Collective,Damiano2016}. Within the stroboscopic approximation, the rapidly oscillating temporal dependence of the Faraday wave is averaged over one forcing cycle, yielding the effective spatial kernel
\begin{equation}\label{eq: strobe}
    W(r)=A_0J_0(k_Fr)\,
\operatorname{sech}\!\left(\frac{r}{l_d}\right),
\end{equation}

where $J_0$ is the zeroth-order Bessel function, $k_F=2\pi/\lambda_F$ is the Faraday wavenumber, $A_0$ is the wave amplitude, and $l_d$ is the characteristic decay length of the wavefield. This functional form captures the essential features of the experimentally measured wavefield~\citep{Damiano2016} and provides a simple framework for understanding how wave interference generates effective interactions between droplets. 

For two droplets separated by a distance $r$, their wavefields superimpose linearly, generating an effective interaction determined by the local wavefield gradient. The resulting radial interaction force is
\[
F(r) = A_0 \operatorname{sech}\!\left(\frac{r}{l_d}\right)
\left[
k_F J_1(k_F r) + \frac{1}{l_d} J_0(k_F r)\tanh\!\left(\frac{r}{l_d}\right)
\right],
\]
where the sign of $F(r)$ reflects the local wavefield gradient: an inward gradient ($F<0$) drives droplets together, whereas an outward gradient ($F>0$) pushes them apart. Figure~\ref{Fig2}(a) shows the corresponding interaction profile for the experimental parameters ($\lambda_F\approx4.75$\,mm and $l_d\approx1.5\,\lambda_F$). The zero crossing defines a characteristic equilibrium separation around which neighbouring droplets organize.

The interaction landscape evolves systematically with droplet size (Fig.~\ref{Fig2}b). Small walker-sized droplets ($R\approx0.4$\,mm) experience predominantly repulsive interactions at short range, while intermediate superwalkers ($R\approx1.1$\,mm) exhibit only weak attraction. In contrast, larger droplets ($R\approx1.7$\,mm) develop a pronounced attractive interaction that, together with excluded-volume repulsion, establishes a stable equilibrium separation and supports the formation of persistent active clusters. Larger droplets may also generate capillary or hydrostatic interactions through their appreciable interface deformation. Since the droplets undergo large periodic shape changes throughout each forcing cycle, the quantitative contribution of these effects is not straightforward to isolate. We therefore examine whether a wave-mediated interaction framework alone can reproduce the principal experimental observations.

Figure~\ref{Fig2}(c) illustrates the wavefield landscape generated by three large droplets, showing how the superposition principle naturally extends to multi-droplet configurations. The corresponding experimental wavefield is shown in Fig.~\ref{Fig2}(d) for comparison. Because the wavefields superpose linearly, neighbouring interactions coexist while preserving their equilibrium separations, enabling progressively larger active clusters to form.

The persistent wave-mediated interaction described above behaves as a continuously maintained bond between neighbouring droplets, and we therefore refer to it as a \emph{dynamical bond}. Unlike a static attractive potential, the bond is sustained by the evolving wavefield and possesses three defining characteristics. First, it has a finite equilibrium separation, defined by the zero crossing of $F(r)$ (see Fig.~\ref{Fig2}a), that establishes a stable spacing. Second, following perturbations, the wavefield restores displaced droplets to this equilibrium configuration (see Movies~4--6). Third, the interaction is modular: multiple dynamical bonds coexist within the same active cluster, enabling additional droplets to be incorporated without disrupting the underlying structure (see Movie~3).

\subsection*{\textbf{Morphologies of Active Clusters}}

\begin{figure*}[ht]
  \centering
  \includegraphics[width=\linewidth]{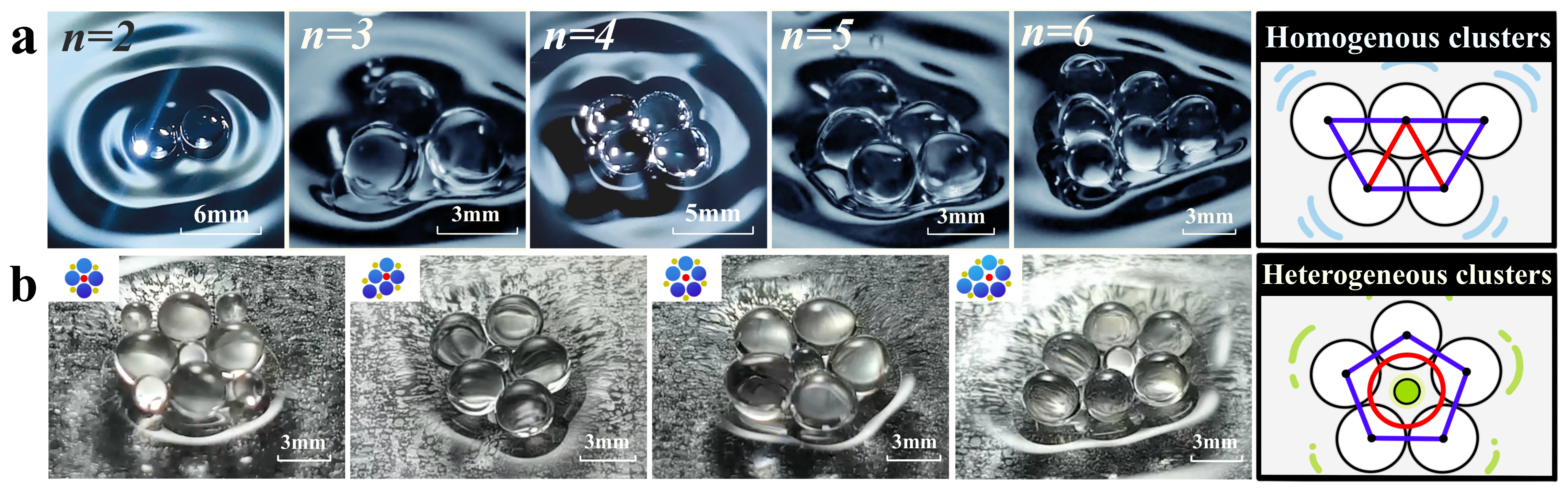}
 \caption{\textbf{Morphologies of active clusters.}
\textbf{(a)} Homogeneous clusters formed from identical large droplets ($N=2$--$6$). Strong wave-mediated interactions produce compact triangular packings stabilized by rigid bonds. Movie~2 provide slow-motion visualizations of these active clusters.
\textbf{(b)} Heterogeneous active clusters containing both large and small droplets. Small droplets mediate coordinating bonds that stabilize cluster geometries beyond triangular packing, including centrally coordinated square configurations. Additional small droplets can also occupy neighbouring wavefield minima around a stable cluster without disrupting its underlying structure (see Supplementary Figure~1).}
  \label{Fig3}
\end{figure*}

The morphology of the active clusters reflects the geometry of the underlying wave-mediated interactions. For large droplets ($R\gtrsim1.2$\,mm), the interaction is dominated by strong radial confinement, establishing a robust equilibrium separation while allowing little variation in bond length or orientation. Consequently, clusters containing $N=2$--$6$ droplets spontaneously adopt compact triangular packings (Fig.~\ref{Fig3}a), the natural consequence of maintaining equal separations in the absence of preferred bond angles.

\begin{figure}[h]
  \centering
  \includegraphics[width=\linewidth]{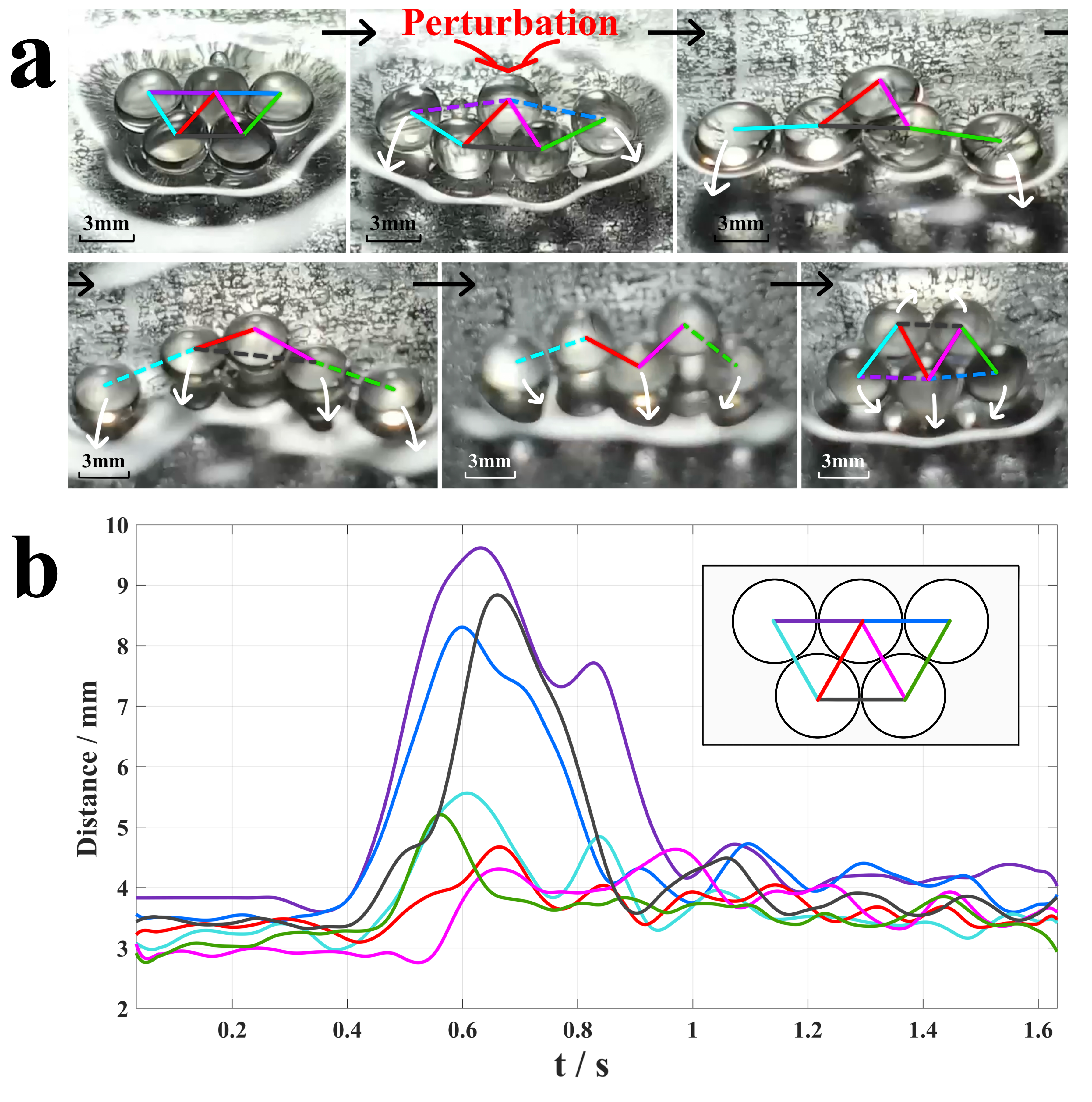}
 \caption{\textbf{Robustness of dynamical bonds.}
\textbf{(a)} Recovery of a five-droplet cluster following a perturbation. The displaced droplet returns to its equilibrium position and the bond is re-established through the wave-mediated interaction (see Movies~4--6).
\textbf{(b)} Evolution of the bond length during the recovery process shown in (a). Following the perturbation, the bond length deviates and then relaxes back to its equilibrium value.}
  \label{Fig4}
\end{figure}

Compact active clusters remain stable up to $N=6$. Beyond this size, we do not observe stable compact configurations. Instead, larger assemblies preferentially adopt chain-like geometries, which become increasingly susceptible to perturbations as their length increases (see Movie~7). In contrast, the clusters with $N\leq6$ remains remarkably robust, repeatedly recovering its compact triangular packing following substantial disturbances (see Movie~8). Figure~\ref{Fig4}a further illustrates the robustness of the wave-mediated interactions following a local perturbation. Displacing a single droplet within a five-droplet cluster temporarily disrupts the associated interaction, but the evolving wavefield drives the droplet back to its equilibrium position, restoring the original separation without external intervention. Figure~\ref{Fig4}b quantifies this recovery by tracking the corresponding bond length, which exhibits a sharp transient increase before relaxing back to its equilibrium value.

Introducing smaller droplets ($R\approx0.5$\,mm) qualitatively changes the wave-mediated interactions. Rather than acting as a rigid connection between neighbouring large droplets, a small droplet responds to the combined wavefield generated by multiple neighbours, modifying the local interaction landscape. This alters the equilibrium arrangement of the surrounding large droplets beyond the triangular packing observed in homogeneous clusters. For example, placing a small droplet at the centre of four large droplets arranged in a rhombus stabilizes a square configuration (Fig.~\ref{Fig3}b; Movie~9). Similar centrally mediated interactions stabilize higher-order cluster geometries that are not observed in assemblies composed solely of identical large droplets.

\subsection*{\textbf{Wave-mediated collective motion}}

\begin{figure*}[h]
  \centering
  \includegraphics[width=0.85\linewidth]{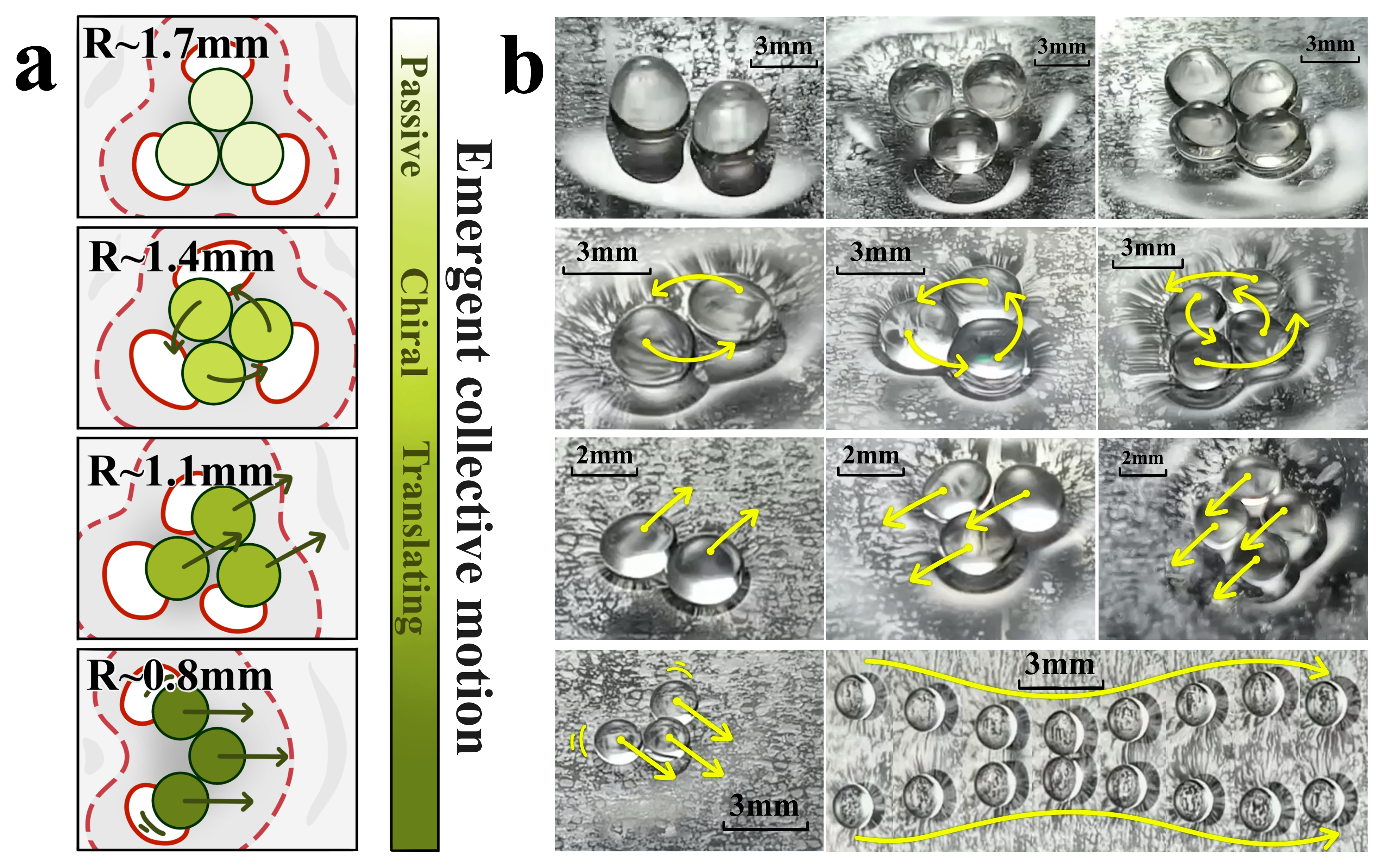}
\caption{\textbf{Collective dynamical states of active clusters.}
\textbf{(a)} Schematic illustrating the qualitative evolution of the wave-mediated interaction with decreasing droplet size. Strong confinement induced by large droplets produces stationary clusters, while weaker confinement from smaller droplets supports rotating and subsequently translating collective states.
\textbf{(b)} Experimental realizations of the corresponding dynamical states for clusters containing two to four droplets. As droplet size decreases, stationary clusters evolve into rotating and translating active clusters, followed by the diverse weakly bound states characteristic of the conventional superwalker regime (see Movies~10--16).}
  \label{Fig5}
\end{figure*}

The same wave-mediated interactions that assemble an active cluster also govern its subsequent motion. Because the interaction is continuously refreshed by the evolving Faraday wavefield, the forces acting on each droplet depend not only on the instantaneous cluster geometry but also on its recent dynamical history. The active cluster therefore behaves as a coherent dynamical object: the droplets remain bound while the entire cluster rotates, translates or undergoes more complex collective motion.

The collective dynamics of the active clusters are governed by the interplay between droplet size and the persistence of the underlying Faraday wavefield. Each bouncing droplet continuously generates waves that survive over many forcing cycles, so that the local wavefield reflects both the instantaneous cluster configuration and its recent evolution. The persistence of this wavefield is controlled by the driving amplitude through the wave memory~\citep{Eddi2011}, while droplet size sets the characteristic interaction geometry and equilibrium separation within the cluster. Together, these parameters shape the wave-mediated interaction landscape experienced by the droplets. In multi-droplet clusters, the delayed wavefield generates force components that are not purely radial but also possess tangential components. Whereas the radial components maintain the cluster geometry, the tangential components act collectively on the droplets, driving coherent rotation or translation while preserving the underlying bonded structure.

Figure~\ref{Fig5}b summarizes the collective dynamical states observed for active clusters containing two to four droplets. As the droplet size is reduced, the same wave-mediated interaction supports a progression of increasingly mobile cluster states while maintaining the underlying bonded structure. Large droplets form stationary clusters, reflecting the strong confinement imposed by the wavefield. With decreasing droplet size, this confinement weakens and persistent rotational motion emerges while the cluster geometry remains intact (Movie~10). A further reduction in size produces translating active clusters that move collectively across the bath without losing their structural coherence (Movie~11).

At droplet sizes characteristic of conventional superwalkers ($R\lesssim1.1$\,mm), an even richer family of weakly bound states appears, consistent with previous observations~\citep{Valani2019Superwalking}. Two-droplet systems exhibit chasing (Movie~12), orbiting (Movie~13), promenading (Movie~14), and outward spiralling states (Movie~15), while three-droplet clusters robustly adopt a plane-like walking configuration with two droplets leading and one trailing (Movie~16). Together, these observations demonstrate that modest changes in the wave-mediated interaction are sufficient to generate a diverse spectrum of coherent collective dynamics, spanning stationary, rotating, translating and weakly bound active states.

\subsection*{\textbf{Chiral Active Clusters}}

\begin{figure*}[h]
  \centering
  \includegraphics[width=\linewidth]{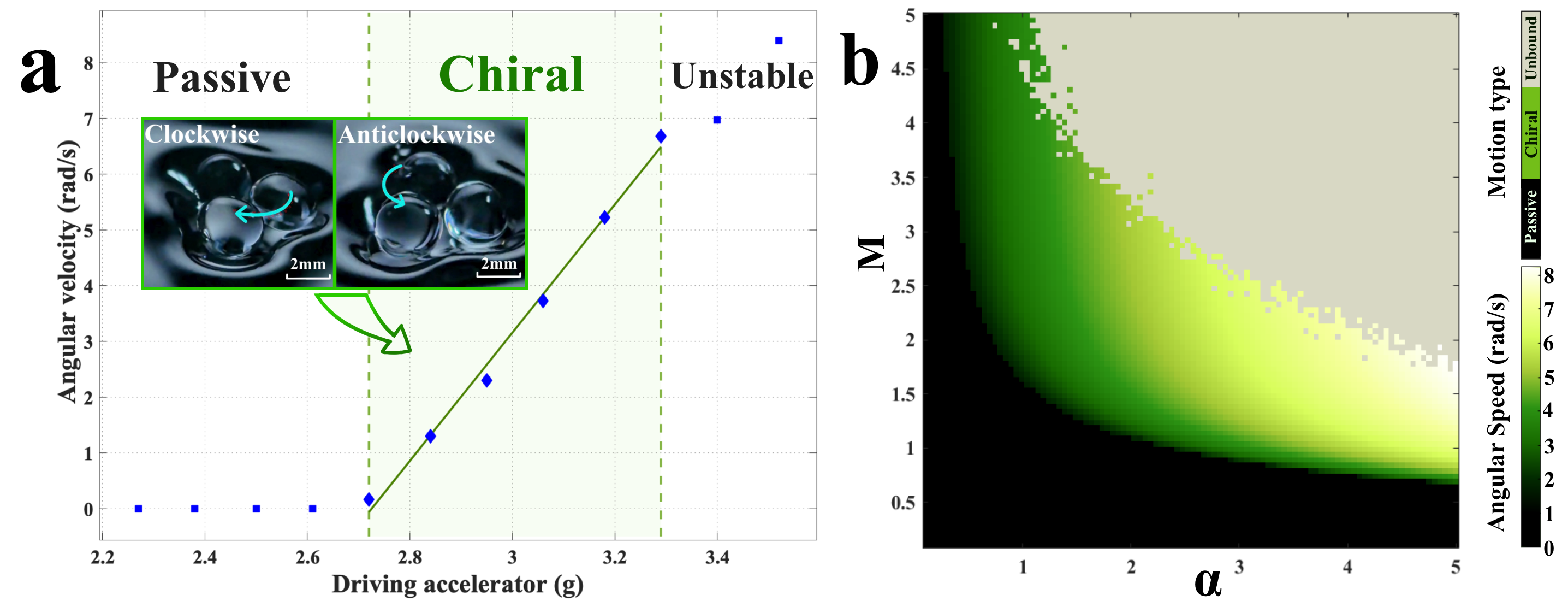}
 \caption{\textbf{Chiral rotation of active clusters.}
\textbf{(a)} Rotating three-droplet active cluster. Three identical droplets of radius $R\approx 1.5$\,mm assemble into a triangular cluster that undergoes persistent clockwise or anticlockwise rotation with equal probability (inset; Movie~17). The main panel shows the measured angular velocity as a function of driving acceleration (blue markers). Above a threshold, persistent rotation emerges and the angular velocity increases approximately linearly (blue diamonds) over the experimentally accessible range (green line: linear fit with $R^2=0.994$), while the cluster geometry remains unchanged. At sufficiently large driving accelerations (blue squares), the rotating state loses stability (Movie~18).
\textbf{(b)} Phase diagram predicted by the stroboscopic model in the $(\alpha,M)$ parameter space. Three dynamical regimes are identified: stationary, rotating, and unstable. Within the rotating regime, the colour scale indicates the predicted angular velocity, which increases monotonically with memory parameter ($M$). The model captures the experimentally observed transition to persistent rotation and yields rotation rates that are consistent with experiment.}
  \label{Fig6}
\end{figure*}

Among the dynamical states described above, the rotating three-droplet cluster provides the simplest and most robust example of a coherent active cluster. Once formed, the cluster undergoes persistent chiral rotation while maintaining its underlying triangular geometry. We use this state to examine how the collective dynamics evolve with the driving conditions.

Three identical droplets spontaneously assemble into a triangular cluster that undergoes persistent rotation, selecting either clockwise or anticlockwise motion with equal probability (Fig.~\ref{Fig6}a, inset; Movie~17). Throughout the motion, the droplets preserve their triangular geometry, so that the cluster behaves as a coherent chiral active object. The rotation rate depends continuously on the driving acceleration. Below a threshold the cluster remains stationary, whereas above this threshold persistent rotation emerges and the angular velocity increases approximately linearly over a broad range of driving accelerations while the cluster structure remains unchanged (Fig.~\ref{Fig6}a). At sufficiently large accelerations, the rotating state loses stability and the cluster eventually coalesces (Movie~18).

To examine whether the observed collective dynamics can be reproduced by wave-mediated interactions alone, we extend the standard stroboscopic model~\citep{Oza2013Trajectory} widely used for walking and superwalking droplets to the regime of our large droplets (see Methods). High-speed videos show that both conventional superwalkers and the large droplets studied here generate qualitatively similar wavefields (Movie~19), supporting the applicability of the same modeling framework. The horizontal dynamics of droplet \(i\) are governed by
\[
m \ddot{\mathbf{x}}_i + D \dot{\mathbf{x}}_i
=
-mg\nabla h(\mathbf{x}_i,t)
+\sum_{j\neq i}\mathbf{F}^{\mathrm{rep}}_{ij},
\]
where \(m\) is the droplet mass, \(D\) an effective drag coefficient, \(g\) the gravitational acceleration, and \(\mathbf{F}^{\mathrm{rep}}_{ij}\) a short-range repulsive force between dropletas $i$ and $j$, accounting for the finite droplet size. The wavefield \(h(\mathbf{x},t)\) is constructed as the superposition of waves generated by all droplets along their previous trajectories,
\[
h(\mathbf{x},t)
=
\frac{A}{T_F}
\sum_j
\int_{-\infty}^{t}
W\!\left(|\mathbf{x}-\mathbf{x}_j(s)|\right)
e^{-(t-s)/(T_FMe)}
\,\mathrm{d}s,
\]
where \(W(|\mathbf{x}|)\) is the axisymmetric strobed droplet wavefield, \(T_F\) is the Faraday period, and \(Me\) is the wave-memory parameter~\citep{Oza2013Trajectory}. Upon nondimensionalizing lengths by \(k_F^{-1}\) and time by \(m/D\), the dynamics are described by two dimensionless parameters: the wave amplitude \(\alpha\) and the memory parameter \(M\).

Figure~\ref{Fig6}b shows the resulting phase diagram for droplets of radius \(R=1.7\) mm, using a slightly larger effective radius than the measured lift-off radius (\(R\approx1.5\) mm) to account for droplet flattening during impact. Three dynamical regimes emerge. At low wave amplitude and memory, the droplets remain locked in a stationary triangular configuration. Increasing either parameter produces a transition to persistent chiral rotation, while sufficiently large values lead to an unstable regime in which the cluster undergoes irregular motion or eventually disintegrates. Within the rotating regime, the predicted angular velocity increases monotonically with memory, consistent with the experimentally observed increase in rotation rate with driving acceleration (Fig.~\ref{Fig6}a). 

The wave-mediated model reproduces the principal experimental observations, indicating that wave-mediated interactions are sufficient to account for the observed behaviour within the present framework. While additional capillary or hydrostatic interactions may contribute quantitatively, particularly for the largest droplets, they are not required to reproduce the observed phenomenology.

\section*{\textbf{Outlook}}

We have identified a new regime of wave-mediated active clustering in which droplets that are unstable in isolation become persistent active clusters through a shared Faraday wavefield. These clusters exhibit diverse collective dynamical states, including stationary, rotating and translating motion, while remaining coherent. The same wave-mediated interactions that assemble the clusters also sustain their subsequent dynamics, linking cluster morphology directly to collective motion.

Beyond the specific droplet system, our experiments illustrate a broader route by which collective interactions generate emergent dynamical objects. The active clusters behave as coherent entities whose properties are not inherited from individual droplets but arise through continual feedback with the surrounding wavefield. In this sense, the cluster itself becomes the fundamental dynamical object, rather than simply a collection of interacting particles.

Shared persistent fields arise naturally in many nonequilibrium systems, including chemical~\cite{Hokmabad2022Chemotactic,Kumar2024}, acoustic~\cite{Ziepke2025Acoustic}, electrical~\cite{Wang2020} and optical active matter~~\cite{doi:10.1126/science.1230020,Wang2021Robot}. The present work therefore suggests that similar mechanisms of active clustering may emerge wherever particles continuously generate and respond to an evolving interaction field. More broadly, it highlights how interactions mediated by a shared field can themselves become the origin of new active entities, rather than simply organizing existing particles.

\bibliographystyle{unsrtnat}
\bibliography{newreferences}

\section*{\textbf{Methods}}

\subsection*{\textbf{Experimental setup}}

Experiments were performed on a circular bath of diameter 120\,mm and fluid depth 10\,mm, filled with silicone oil of kinematic viscosity 20\,cSt (density $\rho \approx 950\,\mathrm{kg\,m^{-3}}$). The bath was mounted on a subwoofer-driven shaking system and driven by a dual-frequency vertical forcing signal consisting of 80\,Hz at 100\% amplitude and 40\,Hz at 25\% amplitude. The waveform was generated using a RIGOL DG1032Z function generator and amplified by a SA-PA003 power amplifier. The peak vertical acceleration $\gamma$ was measured using an accelerometer attached to the bath and calibrated relative to the Faraday instability threshold of $\gamma_F\approx 4.2\,g$ where $g$ is the gravitational acceleration.

Droplets were manually deposited using a syringe with a 2\,mm nozzle. Three characteristic size classes were studied: large droplets ($R>1.2$\,mm), ordinary superwalking droplets ($R\approx0.7$-$1.1$\,mm), and smaller mediator droplets ($R\approx0.5$\,mm). Droplet radii were measured from high-speed images during airborne phases when droplets are approximately spherical.

Horizontal motion was recorded from above at $1200$ frames per second using a Highspeed Vision 300c-U3 camera. Back-illumination from an LED panel provided high contrast for centroid tracking. Droplet positions were extracted using custom image-processing routines in MATLAB.

\subsection*{\textbf{Simulation}}

Numerical simulations were implemented in MATLAB based on the stroboscopic model for walking droplets~\citep{Oza2013Trajectory}. The horizontal motion of droplet \(i\) is governed by
\[
m\ddot{\mathbf{x}}_i + D\dot{\mathbf{x}}_i
=
\mathbf{F}^{\mathrm{int}}_{ii}
+
\sum_{j\neq i} \left(
\mathbf{F}^{\mathrm{int}}_{ij}
+
\mathbf{F}^{\mathrm{rep}}_{ij}
\right),
\]
where \(\mathbf{F}^{\mathrm{int}}_{ii}\) is the self-interaction from the droplet's own wavefield, \(\mathbf{F}^{\mathrm{int}}_{ij}\) is the wave-mediated interaction with other droplets, and \(\mathbf{F}^{\mathrm{rep}}_{ij}\) is a short-range repulsive force accounting for finite droplet size~\citep{4cgg-hnyh}. 

%Our simulation includes only wave-mediated forces and excluded-volume repulsion; no capillary or hydrostatic terms are introduced. Despite this, it quantitatively reproduces the experimental equilibrium separations, cluster geometries, the $N=6$ stability boundary, and—critically—the phase diagram and rotation rates of the chiral clusters (Fig.~\ref{Fig4}b) without fitting to the collective dynamics. This quantitative agreement across multiple independent observables confirms that the wave-only model captures the essential physics.

The wave-mediated interaction force is given by
\begin{align*}
\mathbf{F}^{\mathrm{int}}_{ij}
&=
- \frac{mg A_0 k_F}{T_F}
\int_{-\infty}^{t}
W'\!\left(k_F |\mathbf{x}_i(t)-\mathbf{x}_j(s)|\right)
\frac{\mathbf{x}_i(t)-\mathbf{x}_j(s)}
{|\mathbf{x}_i(t)-\mathbf{x}_j(s)|}
\, e^{-\frac{(t-s)}{T_F Me}} \, \mathrm{d}s,
\end{align*}
which captures the time-retarded forcing from the history of impacts. The short-range repulsion is modeled as
\[
\mathbf{F}^{\mathrm{rep}}_{ij}
=
\bar{K}
\frac{\mathbf{x}_i(t)-\mathbf{x}_j(t)}
{|\mathbf{x}_i(t)-\mathbf{x}_j(t)|}
\left(2R - |\mathbf{x}_i(t)-\mathbf{x}_j(t)|\right),
\]
for \( |\mathbf{x}_i-\mathbf{x}_j| < 2R \), and zero otherwise, where \(2R\) is the droplet diameter.

The drag coefficient is given by~\citep{Oza2013Trajectory}
\[
D = C m g \sqrt{\frac{\rho R}{\sigma}} + 6\pi \mu_a R \left(1+\frac{\rho_a g R}{12 \mu_a f}\right),
\]
where \(C\) is a dimensionless coefficient that depends weakly on system parameters~\citep{Molacek2013}. In the absence of direct calibration for large droplets, we fix \(C=0.2\), a typical value for walkers.

Unless otherwise stated, simulations use parameters corresponding to experiments: forcing frequency \(f=80\) Hz, fluid density \(\rho=950\) kg/m\(^3\), viscosity \(\nu=20\) cSt, surface tension \(\sigma=20.6\times10^{-3}\) N/m, Faraday wavelength \(\lambda_F=4.75\) mm, and Faraday period \(T_F=2/f\). Air viscosity and density are \(\mu_a=1.84\times10^{-5}\) kg/ms and \(\rho_a=1.2\) kg/m\(^3\), respectively. A constant impact phase \(\sin\Phi=0.2\) is chosen, consistent with the value used for walkers~\citep{Oza2013Trajectory}.

We nondimensionalize lengths by \(k_F^{-1}\) and time by \(m/D\), yielding two key parameters: the dimensionless wave amplitude
\[
\alpha = \frac{m^3 g A_0 k_F^2}{D^2 T_F},
\]
and the dimensionless memory
\[
M = \frac{D T_F Me}{m},
\]
where \(Me = T_d/T_F (1-\gamma/\gamma_F)\) is the memory parameter, with a typical decay time \(T_d = 1/54.9\) s.

In dimensionless form, the interaction force becomes
\[
\mathbf{F}^{\mathrm{int}}_{ij}
=
- \alpha \int_{-\infty}^{t}
W'\!\left(|\mathbf{x}_i(t) - \mathbf{x}_j(s)|\right)
\frac{\mathbf{x}_i(t) - \mathbf{x}_j(s)}
{|\mathbf{x}_i(t) - \mathbf{x}_j(s)|}
\, e^{-(t-s)/M} \, \mathrm{d}s,
\]
with wave kernel
\[
W(r) = J_0(r)\,\mathrm{sech}\!\left(\frac{r}{L}\right),
\]
where \(L = l_d k_F\) is the dimensionless spatial decay length, chosen as \(l_d = 1.5\times(2\pi/k_F)\)~\citep{Thompson2020Collective}. The exponential kernel encodes finite memory, such that recent impacts contribute more strongly than distant ones. We also fix the dimensionless spring constant $K=\tilde{K}m/D^2=100$ for short range repulsion.

Time integration was performed using a modified Euler method with time step \(\Delta t = 0.01\) as detailed in \citep{4cgg-hnyh}. The memory integral was evaluated using a discrete history of past positions, truncated when contributions fall below \(e^{-20}\sim10^{-9}\).

\section*{\textbf{Supplementary Information}}

Movies~1-19 and Supplementary Figure~1 are available online. %Supplementary Videos S1-S6 show stroboscopic visualization of $2$-$6$ droplet clusters over one complete vibration cycle, demonstrating that all droplets remain phase-locked to the bath oscillation, justifying the stroboscopic approximation used in the simulations.

\section*{\textbf{Acknowledgements}}
Y. Li acknowledges the support of the National Demonstration Center for Experimental Physics Education, Nanjing University and the School of Physics, Nanjing University.

R.V. acknowledges the support of the Leverhulme Trust [Grant No. LIP-2020-014] and the ERC Advanced Grant ActBio (funded as UKRI Frontier Research Grant EP/Y033981/1).

\end{document}